\documentclass[aps,prl,showpacs]{revtex4}

\usepackage{amssymb,amsmath,amsbsy}

\newcommand{\be}{\begin{equation}}
\newcommand{\ee}{\end{equation}}

\begin{document}
\large

\title{Group theory, coherent states, and the N-dimensional oscillator}

\author{C. R. Hagen\footnote{mail to hagen@pas.rochester.edu} }
\affiliation{Department of Physics and Astronomy\\
University of Rochester\\
Rochester, N.Y. 14627-0171}

\begin{abstract}
The isotropic  harmonic oscillator in $N$ dimensions is shown to have an underlying symmetry group O(2,1)$\times$O($N$) which implies a unique result for the energy spectrum of the system.  Raising and lowering operators analogous to those of the one-dimensional oscillator are given for each value of the angular momentum parameter.  This allows the construction of an infinite number of coherent states to be carried out.  In the $N=1$ case there is a twofold family of coherent states, a particular linear combination of which coincides with the single set already well known for that case.  Wave functions are readily derived which require only the solution of a first order differential equation, an attribute generally characteristic of group theoretical approaches.   
      
  \end{abstract}



\maketitle


\section{I Introduction}

The importance of the simple harmonic oscillator to the study of quantum mechanical systems is virtually impossible to overstate.  Its solution is well known both from the Schr\"{o}dinger equation approach as well as the more elegant algebraic method based on raising and lowering operators.  The extension to the case of $N$ dimensions is quite trivial when formulated in terms of Cartesian coordinates since there is a total separation of variables in that case.  The eigenvalues (eigenfunctions) are thus seen to be simply the sum (product) over the $N$ individual oscillators.  Somewhat less obvious is the analysis when spherical coordinates are employed.  The usual route lies through the separation of radial and angular coordinates and a straightforward Schr\"{o}dinger equation approach yields the energy eigenvalues as well as the wave functions.  The latter are found to be expressible in term of Laguerre polynomials and generalized spherical harmonics.  By contrast the goal here is to derive a solution which is entirely algebraic.  Aside from the aesthetic appeal of such a representation independent approach, it will be seen to imply the existence of coherent states beyond those which have been so widely employed in quantum optics.  

The starting point of this study is the Hamiltonian of the $N$-dimensional isotropic oscillator as given by
$$H={{\bf p}^2\over 2M} +{1\over 2}M{\omega}^2 {\bf x}^2$$
where $p_i$ and $x_i$ with $i=1,2,...N$  are respectively the canonical momenta and coordinates.  These satisfy the commutation relations
$$[p_i,x_j]=-i\delta_{ij}$$
with $\hbar=1$.  The rescalings $x_i=Q_i(\sqrt{ M\omega})^{-1},   p_i=P_i\sqrt{M\omega},$
and the definition $J_3=H/(2\omega)$ yields 
$$J_3={1\over 4}({\bf P}^2+{\bf Q}^2).$$
Rotational invariance is invoked by defining the 
${1\over 2}N(N-1)$ 
angular momentum  operators
 $L_{ij}=Q_iP_j-Q_jP_i$ and the total angular momentum squared
 $L^2={1\over 2}L_{ij}L_{ij}.$
 In keeping with the avowed aim of this work to present the $N$-dimensional oscillator in a strictly group  
 theoretical framework the reader is referred to \cite{ Friedmann} for the angular momentum aspect of the system.
 Using group theoretical methods the eigenvalues of $L^2$ have been shown there to be $\ell(\ell+N-2)$  (a well known result) where $\ell$ is a nonnegative integer. Upon invoking the decomposition 
$$L^2/{\bf Q}^2={\bf P}^2-
{\bf P}\cdot {\bf Q}{1\over {\bf Q}^2}{\bf Q} \cdot  {\bf P}$$
it follows that for a fixed $L^2$ eigenvalue $J_3$ takes the form
$$J_3={1\over 4}({\bf P}\cdot {\bf Q}{1\over {\bf Q}^2}{\bf Q}\cdot {\bf P}+{\ell (\ell +N-2)\over {\bf Q}^2}+{\bf Q}^2).$$

In the following section it is shown that the operator $J_3$ is an element of the algebra of the noncompact group O(2,1) which then readily leads to a determination of the energy spectrum of the $N$-dimensional oscillator.  Worth noting is the fact that this group has previously been used by the author \cite{Hagen} in the $N=2$ case
to determine the phase shifts of the Aharonov-Bohm effect.  Section {\bf III} uses this algebra to construct a one parameter coherent state for each $\ell$ value.  The wave functions of the system are subsequently derived in {\bf IV} by direct application of the O(2,1) framework.  The exceptional case $N$=1 is discussed in section {\bf V} with connection being made  to well known coherent state results.  Section {\bf VI} summarizes some of the main results.    

 \section{II O(2,1) algebra of the N-dimensional  oscillator}
 
 The demonstration of a O(2,1) algebra requires the introduction of two additional operators, specifically
 $$K_1=-{1\over 4}[{\bf P}\cdot {\bf Q}{1\over {\bf Q}^2}{\bf Q}\cdot {\bf P}+ {\ell (\ell +N-2)\over {\bf Q}^2}-{\bf Q}^2]$$
 and
 $$K_2=-{1\over 4}({\bf P}\cdot {\bf Q}+{\bf Q}\cdot {\bf P}).$$
 The set $J_3, K_1,$ and $K_2$ is closed under commutation; specifically
 $$[J_3, K_1]=iK_2$$
 $$[J_3, K_2]=-iK_1$$ 
 $$[K_1, K_2]=-iJ_3$$
 and thus provides a realization of the algebra of O(2,1).  Since the operators $L_{ij}$ all commute with the set $J_3$,  $K_1$, and $K_2$, one has invariance with respect
 to the direct product O($N)\times$O(2,1).  
 The Casimir operator of the O(2,1) part of this algebra is defined by
 $$C=J_3^2-K_1^2-K_2^2$$
 and is found by direct calculation to be given by 
 $$C={1\over 4}[(\ell +{N\over 2}-1)^2-1].$$
 
 It is convenient to define operators $K_+$ and $K_-$ as given by
 $$K_{\pm}\equiv K_1\pm iK_2.$$
 These respectively raise and lower by one unit the eigenvalues of $J_3$ as is evident from the commutation relations
 $$[J_3, K_{\pm}]=\pm K_{\pm}.$$
 Since repeated action of $K_-$ on any eigenket of $J_3$ yields states of increasingly lower energy, it is clearly necessary by virtue of the positive definiteness of the Hamiltonian to require that for each value of $\ell$ there must be a state which is annihilated by $K_-$.  In other words there must be some eigenvalue $m_0$ of $J_3$ such that 
 $$K_-|m_0,\ell\rangle=0.$$ 
 This is, of course, a straightforward adaptation of the technique used to solve for the spectrum of the one-dimensional oscillator.
 Rewriting the Casimir operator as 
 $$C=J_3^2-J_3-K_+K_-$$
 is seen to imply that
 $$C|m_0,\ell\rangle=(m_0^2-m_0)|m_0,\ell\rangle.$$
 This yields the quadratic equation 
 $$(m_0-{1\over 2})^2={1\over 4}(\ell +{N\over 2}-1)^2$$
 which immediately gives the required eigenvalue $m_0$ as ${\ell\over 2}+{N\over 4}$ and/or $1-{\ell\over 2}-{N\over 4}$.
 Upon application of $K_+$ to the state $|\ell,m_o\rangle$ a succession of $n$ times ($n$=0,1,2,...) it is readily inferred that the energy spectrum $E_{n,\ell}$ is given by
 $$E_{n,\ell}=\omega(2n+\ell +{N\over 2})$$
 and/or
$$E_{n,\ell}=\omega(2n+2-\ell-{N\over 2}).$$
The former set of eigenenergies is the one which has been obtained in various contexts by separation of variables in the relevant Schr\"{o}dinger equation and also by a partial group theoretical analysis of the 
Schr\"{o}dinger equation \cite{Bacry}.  Thus it is only the second set which requires detailed comment.

Since the energies of the system are necessarily positive, it is only the cases $N\leq 4$ of the second set which are possibly relevant.  For $N=4$ positivity requires  $\ell=0$ and it follows that the energy eigenvalue is zero.  It implies a trivial realization of O(2,1).  Considering $N$=2 one finds that $\ell$ equal zero and one are both possible.  The latter is also a trivial realization of O(2,1) while the former corresponds to an eigenvalue already included in the other set of eigenvalues.  This leaves only $N$=3 and 1 for discussion.  If $N$=3 and $\ell$=0, one finds a ground state of ${1\over 2}\omega$, a value which must be excluded by the Cartesian coordinate solution for $N$=3.  Finally, one is left only with the $N$=1 case and indeed this second set solution requires inclusion as it corresponds to the odd parity solution of the one-dimensional oscillator.  One concludes that only the usual spectral values of the $N$-dimensional oscillator are relevant except for $N$=1.  That special case is considered in section {\bf V} .

Mention should, of course, be made of the fact that the degeneracy of a given energy eigenvalue for all $N\geq 2$ must be independent of whether the system is quantized in Cartesian or spherical coordinates.  That equality is by no means obvious, but is explicitly demonstrated in the Appendix.

\section{III Coherent states of the N-dimensional oscillator}

One seeks to construct eigenstates of the operator $K_-$ which are analogous to the eigenstates of the usual annihilation operator of the $N=1$ case \cite{The literature}.  Specifically, one requires states $|k\rangle$ such that 
$$K_-|k\rangle=k|k\rangle$$
for $k$ an arbitrary complex number.  This requires that the action of $K_{\pm}$ on the energy eigenstates be determined and is accomplished in close analogy to the case of the $a$ and $a^{\dagger}$ operators (${1\over \sqrt{2}}(Q\pm iP)$ respectively) for the one dimensional oscillator.  In particular one requires the evaluation of $\sigma_n^{\pm}$ as defined by 
$$K_{\pm}|n,\ell\rangle=\sigma_n^{\pm} |n\pm1,\ell\rangle.$$   
It is readily shown that 
$$(\sigma_n^{\pm})^2=\langle n,\ell |(J_3^2\pm J_3-C)|n,\ell\rangle$$
and thus
$$\sigma_n^{+}=\sqrt{(n+1)(n+\alpha +1)}$$
and
$$\sigma_n^{-}=\sqrt{n(n+\alpha)}$$
where $\alpha\equiv \ell+{N\over 2}-1$ and an undetermined phase has been absorbed into the definition of the states.  It should be noted that this does not apply to the $N=1$, $J_3={3\over 4}$ case as will be addressed in section {\bf V}.  

This is seen to allow the construction of $K_-$ eigenstates $|k\rangle$ as
$$|k\rangle=\sum_{n=0}^{\infty} k^nB_n K_+^n|0,\ell \rangle$$
for some set $B_n$ to be determined.  Using the result for $\sigma_n^{\pm}$ as given above 
there results 
$$|k\rangle=B_0 |0,\ell\rangle +\sum_{n=1}^{\infty}k^nB_n\prod_{i=1}^{n}[i(i+\alpha)]^{1\over 2}|n,\ell\rangle.$$
When $K_-$ acts on this one obtains
$$K_-|k\rangle=k[B_1(1+\alpha)|0.\ell\rangle+
\sum_{n=1}^{\infty}k^n B_{n+1} \prod_{i=1}^{n}[i(i+\alpha)]^{1\over 2} (n+1)(n+1+\alpha) |n,\ell\rangle]$$ 
so that $|k\rangle$ is an eigenvector of $K_-$ with eigenvalue $k$ provided that 
$$B_n=B_{n+1}(n+1)(n+1+\alpha)$$
and
$$B_0=B_1(1+\alpha).$$
These relations are satisfied with the choice
$$B_n={\Gamma(1+\alpha)\over n!\Gamma(n+1+\alpha)}$$
thereby establishing the existence of coherent states for the $N$-dimensional oscillator.

The time development of these states is obtained in the usual way by considering 
$$|k\rangle_t\equiv\exp(-iHt)|k\rangle$$
readily yielding the result 
$$|k\rangle_t=e^{-i\omega ({\ell+{N\over 2}})t}\sum_{n=0}^{\infty}(ke^{-2i\omega t})^nB_nK_+^n|0,\ell\rangle$$
thereby displaying the fact that the time development aside from a trivial phase factor consists solely in the replacement of $k$ by the time dependent term $ke^{-2i\omega t}.$
This result closely parallels the well known time development of the usual $N=1$ coherent states.
The norms of these states are readily computed from
$$\langle k|k\rangle=\langle 0|\sum_{n=0}^{\infty}k^{\star n} K_-^{n}B_n^{\star} \sum_{n'=0}^{\infty}k^{n'} K_+^{n'}B_{n'}|0\rangle.$$
This reduces to 
$$\langle k|k\rangle=\sum_{n=0}^{\infty}(kk^{\star})^n|B_n|^2{n!\Gamma(n+\alpha+1)\over \Gamma(\alpha +1)}$$
whence it follows that
$$\langle k|k\rangle=\sum_{n=0}^{\infty}(kk^{\star})^n{\Gamma(1+\alpha)\over n!\Gamma(n+1+\alpha)}.$$
More succinctly this is written as 
$$\langle k|k\rangle=\Gamma(1+\alpha)I_{\alpha}(2|k|)|k|^{-\alpha}$$
where $I_{\alpha}$ denotes the usual modified Bessel function.

It is well to remark here that the construction of coherent states for the $O(2,1)$ group has also been considered by Barut and Girardello \cite{Barut} from a purely mathematical perspective.  Because of the fact that the concern here is with a specific physical system (i.e., the $N$-dimensional oscillator), the relevant symmetry group in the current application is O(2,1)$\times$O(N) but only one of the two types of representations obtained in ref. 5 turns out to be physically relevant.  This is in direct analogy to the case of the Aharonov-Bohm effect where again only half of the O(2,1) representations are actual solutions of the physical system being considered \cite{Hagen}.

\section{IV Wave functions of the N-dimensional oscillator}

While the principal focus of this work is the algebraic aspect of the $N$-dimensional oscillator,
it in fact provides a direct route to the wave functions of the system which requires only the solution of first order differential equations.  One begins with the energy eigenvalues $E_{n\ell}=\omega(2n+\ell+{N\over 2})$, leaving to section {\bf V} the exceptional case of $N$=1.  The ground state satisfies the equation 
$$K_-|0,\ell \rangle=0$$
which in the coordinate representation implies
$$\langle r|(J_3-{1\over 2}Q^2-{i\over 2}Q\cdot P-{N\over 4})|0,\ell\rangle=0.$$
Using the fact that the eigenvalue of $J_3$ for this state is ${1\over 2}(\ell+{N\over 2})$ it follows that
$$({1\over 2}\ell-{1\over 2}r{\partial\over {\partial r}}-{1\over 2}r^2)\langle r|0,\ell\rangle=0$$
so that up to a normalization $\psi_{0\ell}(r)\equiv\langle r|0,\ell\rangle$ is given by
$$\psi_{0\ell}(r)=r^{\ell}e^{-r^2/2}L_0^{(\alpha)}(r^2).$$
In writing this equation use has been made of the Laguerre polynomial $L_n^{(\alpha)}(r^2)$ which is a constant for the case $n=0$.  It clearly anticipates the result for general $n$.

Making the ansatz that 
$$\psi_{n\ell}(r)=A_nr^{\ell}e^{-r^2/2}L_n^{(\alpha)}(r^2)$$
the equation $$K_+|n,\ell\rangle=\sigma_n^+  |n+1,\ell\rangle$$
implies
$$-(n+{\ell\over 2}+{N\over 2}-{1\over 2}r^2+{1\over2}r{\partial\over\partial r})A_nr^{\ell}e^{-r^2/2}L_n^{(\alpha)}(r^2)=\sigma_n^+ A_{n+1}r^{\ell}e^{-r^2/2}L_{n+1}^{(\alpha)}(r^2)$$
or with $r^2=x$
$$-(n+\alpha+1-x+x{\partial\over \partial x} )A_nL_n^{(\alpha)}(x)=\sigma_n^+A_{n+1}L_{n+1}^{(\alpha)}(x).$$
The Laguerre polynomials given by 
$$L_n^{(\alpha)}(x)={1\over n!}e^xx^{-\alpha}({\partial ^n\over \partial x^n})e^{-x}x^{n+\alpha}$$
are readily seen to be a solution of this equation for the coefficient choice
$$-(n+1)A_n=\sigma_n^+A_{n+1}.$$
This allows one to write the normalized wave functions as
$$\psi_{n\ell}(r)=(-1)^n[{2n!\over \Gamma(n+\alpha+1)}]^{1\over 2}r^{\ell}e^{-r^2/2}L_n^{(\alpha)}(r^2)$$
in agreement with the usual result \cite{Lynch} save for the innocuous factor of $(-1)^n$.  (The latter emerges from the fact that in the group theoretical approach employed here wave functions corresponding to different $n$ values are related by successive applications of the $K_+$ operator.)
 
 Wave functions of the coherent states are also readily derived.  The result is expressible in the form 
 $$\langle r|k\rangle=r^{-N/2+1}e^{-r^2/2}[2\Gamma(1+\alpha)]^{1\over2}e^{-k}I_{\alpha}(2rk^{1/2})k^{-\alpha/2}.$$
 This yields the asymptotic behavior 
 $$\lim_{r \rightarrow \infty}r^{N-1}|\langle r|k\rangle_t|^2={1\over 2\pi}\Gamma(1+\alpha)\exp[{-(r-2\sqrt{k}\cos\omega t)^2}]e^{2k} k^{-(\alpha+{1\over 2})},$$
 a result which closely resembles the corresponding result obtained in  the $N=1$ case for coherent states.  The comparison is made even more striking by noting that  in the limit $r, k\rightarrow \infty$ with $\sqrt{k}<<r$ one has
 $$\lim_{r,k\rightarrow\infty}{r^{N-1}|\langle r|k\rangle_t|^2\over \langle k|k\rangle}={1\over \sqrt{\pi}}\exp[-(r-2\sqrt{k}\cos\omega t)^2].$$
 This result is in exact agreement with the normalized probability density which is well known to describe coherent states in the $N=1$ case.  It is worth noting that in the prescribed limits there is no longer any residual dependence on the angular momentum parameter $\ell$.
  
 \section{V The $N=1$ Case}
 
 It has been observed in this work that the $N=1$ case is exceptional.  In fact only for $N=1$ are both of the energy eigenvalue sets $E=\omega(2n+{1\over 2})$ (even parity) and $E=\omega(2n+{3\over 2})$ (odd parity)  allowed.
For the former case the parameters $\sigma_n^{\pm}$ are immediate as 
$$\sigma_n^+=\sqrt{(n+1)(n+{1\over 2})}$$
and
$$\sigma_n^-=\sqrt{n(n-{1\over 2})}.$$
For the odd parity case the corresponding results (here called ${\tilde{\sigma}}_n^{\pm}$) are readily found to be 
$${\tilde{\sigma}} _n^{+}=\sqrt{(n+1)(n+{3\over 2})}$$
and
$${\tilde{\sigma}}_n^-=\sqrt{n(n+{1\over 2})}.$$
It follows that the wave functions for these two sets (in terms of the more appropriate coordinate $x$) are proportional to $e^{-x^2/2}L_n^{(-1/2)}(x^2)$ and $xe^{-x^2/2}L_n^{(1/2)}(x^2).$
As these Laguerre polynomials are the Hermite polynomials $H_{2n}(x)$ and $H_{2n+1}(x)$ respectively, one finds  the unified standard result for the wave functions of the one dimensional oscillator
 $$\psi_n(x)=[2^n n!]^{-{1\over 2}}\pi^{-{1\over4}}H_n(x)e^{-x^2/2}.$$
 
 Coherent states for $N=1$ are twofold.  For even parity they follow from those previously derived by specializing to $\ell=0$, $N=1$.  Specifically
 $$|k\rangle=\sum_{n=0}^{\infty}k^n K_+^n{\Gamma({1\over 2})\over n!\Gamma(n+{1\over 2})}|0\rangle.$$
 For the odd parity case the results for ${\tilde{\sigma}}_n^{\pm}$ yield the coherent state
 $$|\tilde{k}\rangle=\sum_{n=0}^{\infty}{\tilde{k}}^n K_+^n{\Gamma({3\over 2})\over n!\Gamma(n+{3\over 2})}|\tilde{0}\rangle$$
 where $\tilde{0}$ denotes the lowest odd parity state.  The wave functions for these states are given by 
 $$\langle x|k\rangle={\pi}^{-{1\over 4}}e^{-x^2/2}e^{-k}\cosh(2x\sqrt{k})$$
 and
$$\langle x|\tilde{k}\rangle={\pi}^{-{1\over 4}}e^{-x^2/2}e^{-{\tilde{k}}}\sinh(2x\sqrt {\tilde{k}})$$
 respectively, and have the norms
 $$\langle k|k\rangle=\Gamma({1\over 2})I_{-{1\over 2}}(2|k|)|k|^{{1\over 2}}$$
 and
 $$\langle \tilde{k}|\tilde{k}\rangle=\Gamma({1\over 2})I_{1\over 2}(2|\tilde{k}|)|\tilde{k}|^{{1\over 2}}.$$
 More simply these are written as $\cosh(2k)$ and $\sinh(2\tilde{k})$ respectively.  The implied time dependent probability densities are readily seen to be given by

\begin{equation*}
\begin{split}
 |\langle x|k\rangle_t|^2  =   {1\over 4}{1\over \sqrt{\pi}}e^{2k}  \{ \exp  [-{1\over 2}(x-2\sqrt{k}\cos\omega t)^2]\exp(-2ix\sqrt{k}\sin\omega t)  \\ 
+  \exp[-{1\over 2}(x+2\sqrt{k}\cos\omega t)^2] \exp(2ix\sqrt{k}\sin\omega t)  \}^2
\end{split}
\end{equation*}

 and
 
 \begin{equation*}
\begin{split}
  \langle x|\tilde{k}\rangle|^2={1\over 4}{1\over \sqrt{\pi}}e^{2\tilde{k}}\{\exp[-{1\over 2}(x-2\sqrt{\tilde{k}}\cos\omega t)^2] \exp(-2ix\sqrt{\tilde{k}}\sin\omega t) \\
  -\exp[-{1\over 2}(x+2\sqrt{\tilde{k}}\cos\omega t)^2\exp(2ix\sqrt{\tilde{k}}\sin\omega t)] \}^2.
  \end{split}
\end{equation*}
These are precisely what one would expect from the usual one-dimensional coherent state approach, given the fact that the development presented here effectively makes a separation into even and odd parity states. 

In the usual formulation of the one-dimensional oscillator there is a single coherent state
$$|c\rangle=\exp(ca^{\dagger})|0\rangle.$$
It includes states of both parities, a fact that suggests the existence of an appropriate combination of the two coherent states found here (each of which contains states of only a given parity) that is equivalent to $|c\rangle$.  Such a construction is facilitated by noting that 
 $$|\tilde{0}\rangle=a^{\dagger}|0\rangle.$$
 Also, since $K^+={1\over 2}(a^{\dagger})^2$ one writes $k=\tilde{k}={1\over 2}c^2$ and seeks a realization of $|c\rangle$ of the form 
 $$|c\rangle=\sum_{n=0}^{\infty}{c^{2n}\over 2^{2n}n!}[(a^{\dagger})^{2n}{\Gamma({1\over2})\over\Gamma(n+{1\over2})}+\lambda (a^{\dagger})^{2n+1}{\Gamma({3\over2}) \over\Gamma(n+{3\over2})}]|0\rangle$$
 with $\lambda$ to be determined.  This relation is in fact verified with the specification $\lambda=c$,
 thereby completing the desired construction of $|c\rangle$ in terms of $|k\rangle$ and $|\tilde{k}\rangle$.  As a final remark it should be noted that the two terms in the above equation are expressible as $\cosh (ca^{\dagger})|0\rangle$ and $\sinh (ca^{\dagger})|0\rangle$.  Neither of these is an eigenvector of $a$ despite the fact that they are individually eigenvectors of $K_-$ by virtue of the fact that the latter is quadratic in $a$.
 
 \section{VI conclusion}
 
 Despite the fact that the algebraic solutions of the one-dimensional and N-dimensional oscillators have long been known, there has not heretofore been a demonstration of the existence of such solutions which incorporate the spherical symmetry of the N-dimensional oscillator.  The construction carried out in this work addresses that shortcoming, providing the first totally group theoretical solution of the N-dimensional oscillator in a spherical basis.  As a natural application of the results obtained it has been seen that wave functions are readily obtained by solving no more than a first order differential equation.  Of most interest perhaps is the construction of coherent states of the N-dimensional oscillator.  These have been shown to yield probability distributions which are asymptotically identical to those which have long been known to characterize the $N=1$ case.  In that latter application the coherent state results obtained have  been shown to reproduce exactly well known results which have been of great significance to the field of quantum optics.  It will be of considerable interest to determine whether there exist domains of comparable applicability in the $N\neq1$ case to other subfields of physics.    
 
 \section{Appendix}
 
 In this appendix it is demonstrated that the degeneracy of a given energy eigenvalue is independent of whether one uses a Cartesian or a spherical basis.  One begins with the well known Cartesian result for the degeneracy $D(\tilde{n})$  $(\tilde{n}=0,1,2,...)$ 
$$D(\tilde{n})={(N+\tilde{n}-1)!\over\tilde{n}!(N-1)!}$$
corresponding to the states of energy $E=\omega(\tilde{n}+{N\over2})$ of the $N$ dimensional isotropic harmonic oscillator. The less familiar result for the multiplicity $d(\ell)$ of the set of angular momentum states labelled by $\ell$ is \cite{Avery} 
 $$d(\ell)={(2\ell+N-2)(\ell+N-3)!\over\ell!(N-2)!}.$$
 To establish the equality of the degeneracies in Cartesian and spherical coordinates it is necessary to show that 
 $$D(\tilde{n})=\sum^{\tilde{n}}_{\ell=0,2,4...}d(\ell)$$
for the case of $\tilde{n}$ being an even number.  In the case of odd $\tilde{n}$ the lower limit is the set $\ell=1,3,5,...$   The proof proceeds  by induction.  Specifically, one assumes that the degeneracies are identical for $\tilde{n}$ and then shows that they agree for $\tilde{n}+2$ as well.  To go from $\tilde{n}$ to $\tilde{n}+2$ one has only to add the single term $d(\ell=\tilde{n}+2)$ to obtain for the spherical degeneracy 
$${N+\tilde{n}-1)!\over\tilde{n}!(N-1)!}+
{(N+\tilde{n}-1)!(2\tilde{n}+N+2)\over(\tilde{n}+2)!(N-2)!}.$$
This can readily be brought to the form 
$${(N+\tilde{n}+1)!\over(\tilde{n}+2)!(N-1)!}$$
thereby yielding $D(\tilde{n}+2).$  The equality of degeneracies for $\tilde{n}$ equal to zero and one is trivially shown and the proof by induction is thereby complete.

\section{Acknowledgment}

Conversations with J. H. Eberly are gratefully acknowledged. 



\begin{thebibliography}{99}

\bibitem{Friedmann}
T. Friedmann and C. R. Hagen, J.  Math. Phys. {\bf 53}, 122102 (2012).

\bibitem{Hagen}
C. R. Hagen, J. Math. Phys. {\bf 54}, 021703 (2013).

\bibitem{Bacry}
H. Bacry and J. L. Richard, J. Math. Phys.  {\bf 8}, 2230 (1967).

\bibitem{The literature}
The literature pertaining to coherent states and their applications to quantum optics is vast.  A useful starting point consists of the books by P. Meystre and M.Sargent, ``Elements of Quantum Optics", Springer-Verlag, (1990) (in particular the references in ch.12) and by W. P. Schleich ``Quantum Optics in Phase Space", Wiley-VCH, (2001)
(in particular the references in ch.11). 

\bibitem{Barut}
A. O. Barut and L. Girardello, Commun. math. Phys. {\bf 21}, 41 (1971).

\bibitem{Lynch}
R. Lynch and H.A. Mavromatis, J.Comp. and App. Math. {\bf 30}, 127 (1990). 

\bibitem{Avery}
J. Avery, ``Hyperspherical Harmonics:  Applications in Quantum Theory", Kluwer Academic Publishers, Dordrecht (1989).


\end{thebibliography}
\end{document}